# Investigation of M1 transitions of the ground-state configuration of In-like Tungsten

W Li[1,2,3], J Xiao[1,2], Z Shi[1,2a], Z Fei[1,2b], R Zhao[1,2c], T Brage[3], S Huldt[4], R Hutton[1,2]* and Y Zou[1,2]*

[1]The Key lab of Applied Ion Beam Physics, Ministry of Education, China
[2]Shanghai EBIT laboratory, Modern physics institute, Fudan University, Shanghai, China
[3]Division of Mathematical Physics, Department of Physics, Lund University, Sweden
[4]Lund Observatory, Lund University, Sweden

Corresponding author: *rhutton@fudan.edu.cn , zouym@fudan.edu.cn

**Abstract.** Three visible lines of M1 transitions from In-like W were recorded using the Shanghai permanent magnet electron beam ion trap. The experimental wavelengths were measured as 493.84±0.15, 226.97±0.13 and 587.63±0.23 nm (vacuum wavelengths). These results are in good agreement with theoretical predictions obtained using large-scale Relativistic Many-Body Perturbation Theory, in the form of the Flexible Atomic Code.

## 1. Introduction

There is a large demand for atomic data of different charge states of Tungsten, since it is considered as a strong candidate for coating material in the International Tokomak Experimental Reactor (ITER), especially in the divertor region. This is due to its excellent thermomechanical properties and its very low erosion under various physical and chemical conditions [1, 2]. Unfortunately, there is very little spectroscopic data available for tungsten in the charge states between $W^{6+}$ and $W^{28+}$ [3], which are important in the divertor region, where the electron temperature will be considerably lower than in the core plasma.

The need for visible spectral lines from tungsten ions in all charge states for ITER diagnostics was pointed out by Skinner in 2006 [4]. Recently we have focused on the charge states of interest in the divertor region of ITER ($W^{6+}$ to $W^{28+}$) [5-10]. For $W^{25+}$, $W^{26+}$ and $W^{27+}$, which are systems with few 4f-electrons (between one and three), most visible lines will originate from magnetic dipole (M1) transitions between fine structure levels of the ground state configurations. $W^{27+}$ has a very simple ground state configuration consisting of only two levels, namely 4f $^2F_{5/2}$ and $^2F_{7/2}$ and just one

M1 transition, which we recently identified and analysed [5]. $W^{26+}$ is more complex having a $4f^2$ ground state configuration. We reported in [6] on our identification of seven transitions between its levels, which made it possible to determine the excitation energy of seven of the thirteen levels in this configuration. The good agreement between theoretical and experimental values for $W^{27+}$ and $W^{26+}$ lends a strong support and confidence to our method for dealing with the few 4f-electron systems. In the work presented here we continue along the lines laid down in the earlier works [5, 6] and study M1 forbidden lines from $W^{25+}$, which is more complex having a ground configuration of $4f^3$ with 41 fine structure levels.

## 2. Experiment

The spectra were recorded using the Shanghai permanent magnet electron beam ion trap (SH-PermEBIT) [11, 12]. Tungsten was injected by the volatile compound $W(CO)_6$ and collided with the electron beam in the drift tube of the EBIT to produce the highly charged ions $W^{q+}$. The photons emitted from the trapped ions were focused by a biconvex lens to be recorded by an Andor SR-303i spectrometer equipped with a 1200 lines/mm grating and an Andor Newton CCD camera (Andor DU940P-BU2). Details of the experimental arrangement can be found in Ref. [5, 6] where we report on forbidden-line spectroscopy of $W^{27+}$ and $W^{26+}$, respectively. Similar to the method employed for $W^{26+}$, the $W^{25+}$ charge state was identified relying on the identification of the $W^{26+}$ and $W^{27+}$ tungsten ions in Ref [5, 6], namely, the first set of lines to appear as the beam energy was lowered from that required to produce the $W^{26+}$ ions were identified as being from the $W^{25+}$ ions. The identifications were also supported by the charge state being in accordance to the ionization potential and all three lines showing the same intensity dependence on the electron beam energy. The wavelength was calibrated by using lines from either a Hg lamp or a Fe hollow cathode lamp or the lines from background carbon and oxygen.

## 3. Calculational method

For $W^{27+}$ and $W^{26+}$ ions, we found excellent agreement between our experimental results and calculations using both Multiconfiguration Dirac-Hartree-Fock method, in the form of the GRASP2K code [13], as well as Relativistic Many-Body Perturbation Theory, utilizing the RMBPT option of the FAC code [14]. It is clear that the RMBPT method is very well-suited for these "isolated" ground state configurations of highly ionized systems and in general this method converges faster than GRASP2K. The identification of these lines was again aided by calculations using both the GRASP2K [13] and the FAC [14] codes, but we focused mainly on the latter and only used GRASP2K to support the identifications.

In the RMBPT approach, the Hilbert space of the full Hamiltonian is partitioned into two sub-spaces, labeled M and N. In the present work, the model space M contains only the $4d^{10}4f^3$ ground configuration where correlation is included to all orders through a configuration interaction (CI) expansion. The N sub-space contains the remaining configurations that are excited by single or double excitations from the M space whose contribution to correlation will be included to second-order perturbation theory. Detailed information about the theoretical method is given in Refs. [14, 15] and can also be found in our previous publications [6, 7, 10].

In order to investigate the convergence of the calculations, we define two maximal principal quantum numbers for the N sub-space: $n_1$ for one-electron excitation and the first electron of the double excitations, and $n_2$ for the second electron of the double excitations. We increase $n_1$ and $n_2$ step-by-step, respectively, to achieve convergence. First, we define $n_2 = 20$ and increase $n_1$ in steps of 10 from 20 to $n_{1max}$, where the results have converged. Second, we set $n_1 = n_{1max}$ and increase $n_2$ in steps of 10 in the range 20 to $n_{2max}$, again with converged results. The maximum orbital angular quantum number was always set to $l_{max} = 15$.

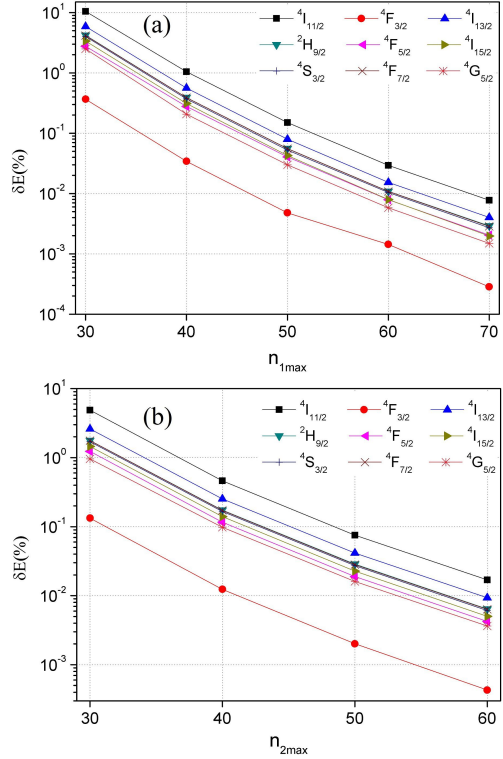

**Figure 1.** Relative convergence of the lowest ten energy levels as the size of maximum principal quantum numbers of the virtual states in the N space (see text). $\delta E$ is the difference in percentages of energy from the previous N space set. (a) $n_{2max} = 20$ and increase $n_{1max}$ in steps of 10 from 20 to 70. (b) $n_{1max} = 70$ and increase $n_{2max}$ in steps of 10 in the range 20 to 60. The y-axis is given in logarithmic scale.

## 4. Results and Discussions

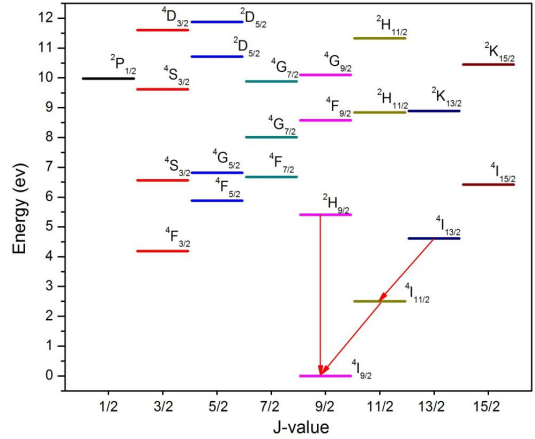

**Figure 2.** A partial energy level diagram of In-like tungsten. The energy levels are taken from our RMBPT calculations. There are 41 energy levels in the ground state configuration of In-like tungsten where the highest levels are close to 28 eV. For convenience we only show levels up to 12 eV as the higher energy levels will not contribute to observable spectral lines. Solid (red) lines represent the M1 emission lines observed in this work

Figure 1 shows the convergence study for the lowest ten excitation energies as a function of $n_{1max}$ and $n_{2max}$, respectively, using the RMBPT method. In the first case, figure 1 (a), it shows the relative convergence with $n_{1max}=70$ the excitation energy is less than 0.01%. For the second set of calculations, figure 1 (b) shows that for $n_{2max}=60$ the convergence is close to 0.01% for $^4I_{11/2}$ and to 0.0003% for $^4F_{3/2}$. The results are therefore well converged and the corresponding partial energy-level diagram is shown in figure 2, where the levels are ordered by their J values and shown with different colors (online). The three M1 emission lines observed in this work are represented by solid (red) lines, see below.

The spectra were recorded for electron beam energies of 730 eV, 780 eV, 830 eV, 880 eV, 1100 eV, and 1180 eV and some of the recorded spectra are shown in figure 3. In these spectra, two lines are marked with arrows (red on-line). The line on the right hand side was recently identified as a $W^{26+}$ line [6,16]. The line at 493.84 ± 0.15 nm, is seen to appear at a lower electron beam energy and just above the ionization potential of $W^{24+}$ and is therefore assumed to be from $W^{25+}$. The experimental wavelength agrees with the prediction from our RMBPT calculation for the $W^{25+}$ ($^4I_{11/2} \rightarrow {}^4I_{9/2}$) line: 492.33 nm. Table 1 shows the comparison of our measured and calculated wavelengths for the three $W^{25+}$ lines studied in this work. The wavelengths of the observed lines are transformed from air to vacuum values using the formula given by Morton [17]. We also compare our data to the calculated results by Radtke et al. using the HULLAC code [18]. The RMBPT calculation results show better agreement with the experimental values where the discrepancy was found to be 0.31%, 0.014% and 0.89%, respectively.

**Table 1.** Computed rates (A), in s$^{-1}$, from present FAC-code and Wavelength (λ), in nm, for three different transitions in W$^{25+}$, from calculations using the FAC (present calculations) and the HULLAC [18] codes and observed in the SH-PermEBIT.

| Transition | A (s$^{-1}$) | λ$_{RMBPT}$ (nm) | λ$_{observed}$ (nm) | λ$_{HULLAC}$ (nm) |
|---|---|---|---|---|
| $^4I_{11/2} \rightarrow {}^4I_{9/2}$ | 283 | 492.33 | 493.84±0.15 | 522.75 |
| $^4I_{13/2} \rightarrow {}^4I_{11/2}$ | 193 | 587.71 | 587.63±0.23 | 583.8 |
| $^2H_{9/2} \rightarrow {}^4I_{9/2}$ | 328 | 229.00 | 226.97±0.13 | 217.51 |

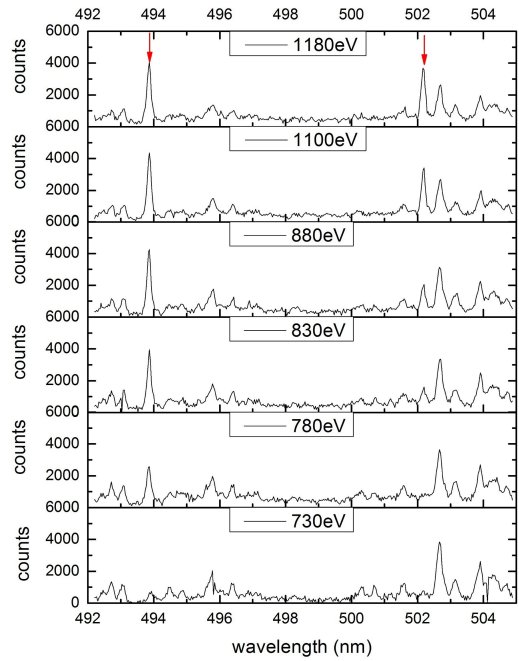

**Figure 3.** Visible spectra of tungsten obtained at the SH-PermEBIT, for different beam energies. The left line at 493.84 nm marked by an arrow is the $^4I_{11/2} \rightarrow {}^4I_{9/2}$ transition in W$^{25+}$ while the right one, also marked by an arrow, is the $^3F_3 \rightarrow {}^3F_2$ transition in W$^{26+}$. For all spectra shown here the beam current was about 5 mA.

As we pointed out above, W$^{25+}$ has a complicated 4f$^3$ ground configuration with the 41 energy levels shown in figure 2. It might therefore at first glance be surprising that not very many lines are observed. There are however three reasons for the lack of observed spectral lines referring to figure 2. First, since there are many levels in this configuration, the excited levels can decay to several different lower levels, leading to possible low intensities in any single branch. Second, the ground state has a high J-value of 9/2 and therefore low-lying levels with total angular momentum of either 3/2 or 5/2 cannot decay to it by M1 transitions. As an example the $^4F_{3/2}$ can only decay radiatively via a high order forbidden transition giving a lifetime in the order of years. This type of level will then decay via collisions, even in very dilute plasma as in an EBIT. In table 2, we give the lowest ten excitation energies and their corresponding computed radiative lifetime. A third reason for the lack of lines in our spectra is that other levels can decay to the lowest J = 9/2 level, but will produce lines outside the range of our spectrometer.

**Table 2.** Excitation energies E, lifetime τ of the lowest ten energy levels of the $4f^3$ ground state configuration of In-like W from our RMBPT calculation. The lifetime of $^4F_{3/2}$ is confirmed by using the MCDHF calculation and big scale Relativistic Configuration Interaction (RCI) calculation.

| Label | E (cm$^{-1}$) | τ (ms) | Label | E(cm$^{-1}$) | τ (ms) |
|---|---|---|---|---|---|
| $^4I_{9/2}$ | 0 | | $^4F_{5/2}$ | 47482 | 18.05 |
| $^4I_{11/2}$ | 20311 | 3.53 | $^4I_{15/2}$ | 51859 | 12.71 |
| $^4F_{3/2}$ | 33737 | 70000 yr | $^4S_{3/2}$ | 53032 | 12.23 |
| $^4I_{13/2}$ | 37326 | 5.18 | $^4F_{7/2}$ | 53990 | 20.15 |
| $^2H_{9/2}$ | 43668 | 2.68 | $^4G_{5/2}$ | 55028 | 19.40 |

## 5. Conclusion

In conclusion, the SH-PermEBIT operates very well in the required electron energy range, and three M1 spectral lines of In-like W have been observed for the first time. The wavelengths were all in excellent agreement (The discrepancies were within 1%.) with our RMBPT result which lends strong support to the identifications. One of these three M1 transitions the $^4I_{11/2} \rightarrow ^4I_{9/2}$, is very strong and may be a good candidate to use in ITER diagnostics in the visible wavelength range.

## Acknowledgements

This work was supported by the Chinese National Fusion Project for ITER No. 2015GB117000, Shanghai Leading Academic Discipline Project No. B107 as well as the Swedish Science Council (Vetenskapsrådet). We would like to thank Professor Gordon Berry for carefully reading the manuscript and for offering very useful comments. We are grateful to the Nordic Center at Fudan University for continued support to our exchange between Fudan and Lund Universities.